\documentclass[aps,pra,preprintnumbers,twocolumn,amsmath,amssymb,showpacs]{revtex4}
\usepackage{graphicx}
\usepackage{amsbsy}
\usepackage{dcolumn}
\usepackage{bm}
\usepackage{hyperref}
\usepackage[latin1]{inputenc}
\newcommand{\rs}{{\bf r}}

\newcommand{\ls}{{\bf l}}
\newcommand{\vs}{{\bf v}}

\newcommand{\ssm}{{\bf s}}
\newcommand{\ns}{{\bf n}}

\newcommand{\rc}{{\bf R}}

\newcommand{\ks}{{\bf k}}

\newcommand{\ps}{{\bf p}}

\newcommand{\ac}{{\bf A}}
\newcommand{\bc}{{\bf B}}
\newcommand{\kc}{{\bf K}}

\newcommand{\qs}{{\bf q}}

\newcommand{\lc}{{\bf L}}

\newcommand{\ssc}{{\bf S}}

\newcommand{\eps}{{\varepsilon}}
\newcommand\ointint{\begingroup
  \displaystyle \unitlength 1pt
  \int\mkern-7.2mu
  \begin{picture}(0,3)
    \put(0,3){\oval(10,8)}
  \end{picture}
  \mkern-7mu\int\endgroup}

\begin{document}
\title{Local phase invariance of the free-particle Schr\"{o}dinger equation in momentum space}
\author{Boyan Obreshkov}
\affiliation{Institute for Nuclear Research and
Nuclear Energy, Bulgarian Academy of Sciences, Tsarigradsko
chaussee 72, Sofia 1784, Bulgaria}
\date{\today}
\begin{abstract}
The local phase-invariance of the momentum-space Schr\"{o}dinger equation 
for free-particle has been used to construct quantum kinematics that describes 
a motion of the particle in external $U(1)$ background gauge field. 
The gauge structure over the momentum space of the particle is interpreted 
in terms of helicity and spin carried by the particle. As a byproduct an effective 
one-particle Schr\"{o}dinger equation of motion for the helicity-carrying particle 
in external potential field is derived. An effect of spin-dependent screening of the external 
potential is predicted, that can affect quantization and splittings of 
energy levels of the particle.

\end{abstract}
\maketitle 

\section{Introduction}
Gauge structure appears in simple dynamic systems \cite{Wilczek}. Observable consequences 
of the gauge structure may be extracted for real systems, 
such as geometrical phase effects \cite{Berry,Wilczek-book}, that can affect the state of a dynamic system
in contact with slowly changing environment. 
For instance, such effects 
have been observed by measuring the changes in the direction of polarization 
of light propagating in a helically coiled optical fiber \cite{fiber}. Observed 
adiabatic rotational energy-level splittings in nuclear quadrupole resonance spectra \cite{NQR} 
has been interpreted in terms of geometrical phase accumulated by nuclear spin states as the 
orientation of the crystal sample is slowly changed.  Quantization of the energy of 
molecular (pseudo)rotation motions by fractional quantum numbers 
has been observed in the light-absorption spectrum of Na$_3$ molecules \cite{Na3}
and explained by an intrinsic geometrical phase affecting the slow 
motion of constituent nuclei. Transport measurements of the 
anomalous Hall resistivity coefficient of ferro-magnetic compounds have been 
related and attributed to the appearance of gauge-structure over 
the momentum space of the charge carriers \cite{Hall1,Hall2}. 
The presence of intrinsic to the material  background monopole-like gauge 
field changes the "normal" direction of propagation of charged particles 
and can lead to observable changes in the transverse conductivity of 
the ferro-magnetic crystal.

The subject of this paper is to study the appearance of gauge structure 
over the momentum space of a free particle, described quantum-mechanically 
by a non-relativistic Schr\"{o}dinger equation of motion.  The paper is organized as 
follows, in Sec. II, the local phase invariance of the free-particle Schr\"{o}dinger equation 
in momentum space is used to introduce a background monopole-like gauge field $\ac(\ps)$, which 
effectively changes the free-particle kinematics without affecting the  
dispersion relation $E=p^2/2$ between energy and momentum. In Sec. III, the appearance of gauge structure over the momentum space 
of the free particle is interpreted in terms of additional two-valued (spin) characteristic
carried by the particle as it moves. In Sec. IV, an effective 
equation of motion for the helicity-carrying particle in external potential field is derived. 
In this approach, the transport of helicity by the particle is expressed 
by a form-factor or path-dependent Berry's phase that has an effect to screen the external 
static background potential field. 
Unless otherwise stated, we use atomic units $(e=\hbar=m_e=1)$. 

\section{Quantum kinematics of a free particle} 
In momentum representation, the free-particle Schr\"{o}dinger equation is 
\begin{equation}
\left(\frac{\ps^2}{2} - E \right) \psi(\ps) = 0, \label{eom}
\end{equation}
where $E$ is the kinetic energy of the particle, and
$\ps$ is the momentum, which is a multiplication
operator. Eq.(\ref{eom}) is invariant under local
change of the phase of the wave-function
\begin{equation}
\psi(\ps) \rightarrow \psi(\ps) e^{-i \Lambda (\ps)}, \label{U1}
\end{equation}
which is because the momentum $\ps$ does not change
\begin{equation}
e^{-i \Lambda (\ps)} \ps e^{i \Lambda (\ps)}=\ps. 
\end{equation}
Therefore the phase of the wave-function is not fixed by Eq.(\ref{eom}). 
Under an infinitesimal local phase transformation of Eq.(\ref{U1}), 
the wave-function changes as 
\begin{equation}
\delta \psi(\ps) = -i \Lambda(\ps) \psi(\ps), 
\end{equation}\
however the derivative 
\begin{equation}
\delta [\nabla_{\ps}  \psi] = -i \Lambda(\ps) \nabla_{\ps} \psi(\ps) -i (\nabla_{\ps} \Lambda(\ps)) \psi(\ps) 
\end{equation}
does not change in the same way. Since the description of the free-particle 
motion is independent on the choice of phase, 
we introduce a derivative that transforms covariantly 
under $U(1)$ phase transformations 
\begin{equation}
{\bf D} = \nabla_{\ps} + i e \ac(\ps) 
\end{equation}
where $e$ is a coupling constant (not necessarily $e=1$), 
and $\ac(\ps)$ is a compensating gauge field, that transforms according to 
\begin{equation}
\ac(\ps) \rightarrow \ac(\ps) + \frac{1}{e} \nabla_{\ps} \Lambda(\ps),  
\end{equation}
and ensures that 
\begin{equation}
{\bf D} \delta \psi = \delta [\nabla_{\ps} \psi] + i e (\delta \ac) \psi + 
i e \ac \delta \psi = -i \Lambda(\ps) {\bf D} \psi(\ps) 
\end{equation}
nothing depends on the arbitrary local phase factor. A gauge-invariant 
one-particle displacement operator $\rc=i {\bf D}$ can be introduced, 
which satisfies the commutation relations 
\begin{equation}
[R_i, p_j]=i \delta_{ij}, \quad [R_i,R_j]= -i e F_{ij}(\ps) 
\end{equation}
where  $F_{ij}$ is anti-symmetric second-rank tensor of the displacement field strength 
\begin{equation}
F_{ij}= \frac{\partial}{\partial p_i} A_j -
\frac{\partial}{\partial p_j} A_i = \eps_{ijk} B_k, 
\end{equation}
$\eps_{ijk}$ is the Levi-Civita symbol and $B_k(\ps)$ labels the components 
of the background magnetic-like field. 
Gauge-invariant angular momentum operator can be introduced $\lc=\rc \times \ps$, however
its components do not satisfy canonical commutation relations
\begin{equation}
[L_i,L_j]=i \eps_{ijk} L_k - i e \eps_{ikl} \eps_{jmn} p_l p_n F_{km}
\end{equation}
and $\{ L_i \}$ are not generators of spatial rotations. The canonical angular momentum 
algebra $[L_i,L_j]=i \eps_{ijk} L_k$ can be restored when the background 
magnetic-like field is rotation-symmetric $\bc(\ps) = B(p) \hat{\ps}$
by the transformation 
\begin{equation}
{\bf L}= \rc \times \ps +e p^2 B(p) \hat{\ps} ,
\end{equation}
The form-factor $B(p)$ can be determined from the requirement that the displacement 
operator transforms as a vector $[L_i,R_j]=i \eps_{ijk} R_k$, which is only satisfied 
when $B(p)=g/p^2$, where $g$ is a field-strength constant. 
The conserved gauge-invariant rotation operator is given by 
\begin{equation}
{\bf L}= \rc \times \ps + e g \hat{\ps} , \label{lcov}
\end{equation}
Eq.(\ref{lcov}) is momentum-space analogue to the angular momentum operator 
$\rs \times (\ps-e \ac(\rs)) - e g \hat{\rs}$ of a charged
particle in an external magnetic field of point monopole charge of strength $g$. 
The rotation symmetry restoration term $e g \hat{\ps}$ is related to the generator of gauge transformations 
of the wave-function $W = eg \hat{\ps} \cdot \ns$  which compensates for the non-symmetric 
response of the gauge-field $\ac(\ps)$ to rotations about the unit-vector $\ns$ \cite{Jackiw3}. 

Analogously, a gauge-invariant extension 
of the generator of Galilei boost transformations can be based on the operators 
\begin{equation}
\kc = \ps t -\rc, 
\end{equation}
where $t$ is the time evolution parameter. The boost operator components do not commute 
\begin{equation}
[K_i,K_j]=[R_i,R_j]=-i e F_{ij}(\ps)
\end{equation}
but are simply related to the conventional generators by change of coordinates 
\begin{equation}
\kc \rightarrow \kc + e \ac = \ps t -\rs \label{Galilei}
\end{equation} 
Under an infinitesimal boost transformation generated by the operators $\{K_i\}$, 
the coordinates $\rc$ change as
\begin{equation}
\delta \rc = \delta \rs - e \delta \ac  = i  [ \delta \vs \cdot \kc, \rc] = \delta \vs t +
\frac{e g}{p^2} \delta \vs \times \hat{\ps} , \label{chang}
\end{equation}
The first term $\delta \vs t$ is the infinitesimal Galilei transformation, which is 
supplemented by a term, which describes an apparent rotation of the 
momentum $\ps$ about the direction of the boost $\delta \vs$. The accompanying 
rotation effect is normally suppressed at high kinetic energies $p^2 \gg eg$. 
The canonical coordinates $\rs$ change in conventional way as $\delta \rs = \delta \vs t$, 
since the variation of the gauge-field $\delta \ac$ can be compensated by re-definition  
of the phase of the wave-function. 

The representation of the modified boost operators in the Hilbert space of states 
is based on exponentials 
\begin{equation}
U(\vs)= \exp(-i \vs \cdot \rc), 
\end{equation}
depending on a velocity vector $\vs$, with the following action onto the wave-function
\begin{equation}
U(\vs) \psi(\ps)= \exp(-i \vs \cdot \rc) \exp(i \vs \cdot \rs) \psi(\ps+\vs). 
\end{equation}
The product of the two exponentials can be expressed by a straight-line integral 
\begin{equation}
\exp(-i \vs \cdot \rc) \exp(i \vs \cdot \rs) = \exp\left( i e \int_{\ps}^{\ps+\vs} d \ks \cdot \ac(\ks) \right) 
\label{Berry-phase}
\end{equation}
that connects the points $\ps$ with $\ps+\vs$. The composition law for the 
generalized boost transformations takes the form 
\begin{equation}
U(\vs_1) U(\vs_2) = \exp[i \omega_2(\ps,\vs_1,\vs_2)] U(\vs_1+\vs_2), 
\end{equation}
where
\begin{equation}
\omega_2(\ps,\vs_1,\vs_2) = e \oint_{\triangle_2}  d \ks \cdot \ac(\ks) 
\end{equation}
is the Berry's phase \cite{Berry}, which is the flux of the 
background magnetic-like field $\bc(\ps)$ through the 
triangle $\triangle_2$ formed by the vertices of the 
momenta $\ps,\ps+\vs_1$ and $\ps+\vs_1+\vs_2$. The phase 
factor violates associativity of the gauge-invariant 
boost transformations (cf. \cite{Jackiw2}), since 
\begin{equation}
[U(\vs_1) U(\vs_2)] U(\vs_3)=e^{i \omega_3(\ps,\vs_1,\vs_2,\vs_3)} 
U(\vs_1) [U(\vs_2)) U(\vs_3)]
\end{equation}
where a three co-cylce phase $\omega_3$ 
\begin{equation}
\omega_3(\ps,\vs_1,\vs_2,\vs_3) = e \ointint_{\Delta_3} \bc \cdot d \ssc
\label{3cocyc}
\end{equation}
is the flux of the background magnetic-like filed through a tetrahedron $\Delta_3$
formed by the vertices of the momenta $\ps$, $\ps+\vs_1$, $\ps+\vs_1+\vs_2$ and $\ps+\vs_1+\vs_2+\vs_3$. 
Associativity can be restored when a Dirac-type quantization condition is satisfied 
\begin{equation}
\omega_3(\ps,\vs_1,\vs_2,\vs_3) = 2 \pi n, 
\end{equation} 
Applying the Stokes theorem to Eq.(\ref{3cocyc}) gives 
\begin{equation}
e \int \int \int d^3 \ks \nabla_{\ks} \cdot \bc(\ks) = 4 \pi e g = 2 \pi n 
\end{equation}
which implies quantization of the product of the two coupling constants $eg=n/2$. 

Since the background magnetic-like field is rotation symmetric,
it can not be written as $\bc = \nabla_{\ps} \times \ac(\ps)$ over the entire momentum space. 
Locally, we can look for a gauge field $\ac(\ps)$ in the form 
\begin{equation}
\ac(\ps)= A(\theta) \nabla_{\ps} \varphi,
\end{equation}
where $(\theta, \varphi)$ are the spherical coordinates of the momentum 
$\ps=(p,\theta,\varphi)$, the equation $\bc = \nabla_{\ps} \times \ac$ is solved by 
\begin{equation}
A(\theta)= -g(1+\cos{\theta}) 
\end{equation}
The gauge field $\ac$ exhibits 
unremovable coordinate-type Dirac string singularity along the line $\theta=0$. 
Singularity-free gauge fields can be defined on two overlapping momentum space patches
\begin{eqnarray}
& & \ac_N= \frac{g}{p} \frac{1-\cos{\theta}}{\sin{\theta}}
\hat{\varphi},
\quad R_N: 0 \le \theta < \frac{\pi}{2}+\eps\nonumber \\
& & \ac_S = -\frac{g}{p} \frac{1+\cos{\theta}}{\sin{\theta}}
\hat{\varphi} \quad R_S: \frac{\pi}{2}-\eps < \theta \le \pi
\end{eqnarray}
where $\ac_N$ is regular on the northern momentum-space
hemi-sphere $R_N$, while $\ac^S$ has support on the southern
hemi-sphere $R_S$. Near the equator $R_N \bigcap R_S$, where the 
gauge-field has a discontinuity, the pair of 
potentials can be related by a gauge transformation
\begin{equation}
\ac_S \rightarrow \ac_S -i e^{-i n \varphi} \nabla_{\ps}
e^{i n \varphi} = \ac_N, 
\end{equation}
Since the gauge-field is not globally defined, the rotation operator $\lc$ is 
not globally defined either. The component of the 
rotation operator onto the space-fixed $z$-axis is two-valued, since 
\begin{equation}
L_z = -i \partial_{\varphi} + eg , \quad (\theta,\varphi) \in R_N 
\end{equation}
or 
\begin{equation}
L_z = -i \partial_{\varphi} - eg , \quad (\theta,\varphi) \in R_S
\end{equation}
depending on the orientation of the wave-vector $\ps$. However, the rotation-symmetric term that restores 
conventional angular momentum algebra 
\begin{equation}
\ssm = eg \hat{\ps}  \label{spin}
\end{equation}
is conserved and does not depend on the momentum space patching. It is related to the 
helicity of the particle by 
\begin{equation}
\lc \cdot \hat{\ps} = \ssm \cdot \hat{\ps} = eg  = n/2 , 
\end{equation}
which is quantized topologically with integer or half integer numbers. 
The free-particle wave-functions of definite helicity $\mu=eg$ are  
eigen-functions of the operators $\lc^2$ and $L_z$. 
The square of the angular momentum operator in Eq.(\ref{lcov}) for the northern 
patch is then given by
\begin{eqnarray}
& & \lc^2=-\frac{1}{\sin^2{\theta}} \left[ \sin{\theta}
\frac{\partial}{\partial \theta} \left( \sin{\theta}
\frac{\partial}{\partial \theta} \right) + \right. \nonumber
\\
& & + \left. \left( \frac{\partial}{\partial \varphi} -i \mu
(1-\cos{\theta}) \right)^2 \right] + \mu^2
\end{eqnarray}
with corresponding rotation operator about the $z$-axis 
$L_z=-i\partial_{\varphi}+\mu$. Angular momentum
eigen-functions can be determined from the equations
\begin{equation}
\lc^2 |l m \mu \rangle = l(l+1) |l m \mu \rangle, \quad
L_z |l m \mu \rangle = m |l m \mu \rangle,
\end{equation}
for $l=|\mu|,|\mu|+1,\ldots$ and $-l \le m \le l$.
Wave-functions are given by sectional (spin-weighted) Wu-Yang monopole harmonics
\begin{equation}
Y_{lm\mu}(\theta,\varphi)=\langle \theta,\varphi |l m \mu
\rangle
\end{equation}
or equivalently expressed by the Jacobi polynomials
$P^{(\alpha,\beta)}_n(z)$
\begin{eqnarray}
& & Y_{lm \mu}(\theta,\varphi)=N_{lm} e^{i (\mu+m)
\varphi} (1-z)^{-(\mu+m)/2}
\times \nonumber
\\
& & \times (1+z)^{-(\mu-m)/2}  P^{(-\mu-m,-\mu+m)}_{l+m}(z) , \label{wu-yang}
\end{eqnarray}
where $z=\cos{\theta}$ and $N_{lm}$ are normalization constants. 
The wave-functions $Y_{lm \mu}$ of half-integer angular momentum $\mu=n/2$
correspond to spinor representations of the rotation group, since they are related to 
the Wigner's rotation functions by
$Y_{l m \mu}(\theta,\varphi)=D^l_{\mu m}(-\varphi,\theta,\varphi)$. The total one-particle 
wave-function, that is an eigen-function of $H,\lc^2, L_z$ is characterized 
by four quantum numbers and given by 
\begin{equation}
\psi_{klm \mu}(\ps)= \frac{\delta(p-k)}{2 k} Y_{lm \mu}(\hat{\ps}), 
\end{equation}
where $k=\sqrt{2 E}$ is a characteristic wave-number. When $\mu=0$, these wave-functions reduce
to the conventional spherical harmonics $Y_{l m}(\theta,\varphi)$. 

\section{Helcitiy and spin of a free particle}
The effects of the gauge-field can be expressed by the non-integrable phase 
factor 
\begin{equation}
\exp \left(i \int_{\ps}^{\qs} d \ks \cdot \ac(\ks) \right)
\end{equation}
which accompanies the translation motion of the free-particle. The gauge-potential 
one-form $A=d \ks \cdot \ac(\ks)$ is singular and can not be defined globally over the two-dimensional 
unit sphere $S^2$. A description that avoids gauge patching of the momentum space can be based 
on constructing a Hopf bundle \cite{monopole-book,Aitch} over the two-dimensional unit sphere $S^2$. We 
further restrict our analysis to the minimal non-trivial helicity quantum numbers $\mu=+1/2$
and $\mu=-1/2$. A regular gauge potential one-form can be defined on the three-dimensional 
unit sphere $S^3$ in four-dimensional Euclidean space $\mathbb{R}^4$. The 3-sphere 
can be parametrized by four coordinates 
\begin{eqnarray}
& & p_1 = \cos\frac{\theta}{2} \cos{\alpha} \nonumber \\
& & p_2 = \cos\frac{\theta}{2} \sin{\alpha} \nonumber \\
& & p_3 = \sin\frac{\theta}{2} \cos(\varphi+\alpha) \nonumber \\ 
& & p_4 = \sin\frac{\theta}{2} \sin(\varphi+\alpha)
\end{eqnarray}
such that $p_1^2+p_2^2+p_3^2+p_4^2=1$. These four coordinates can be grouped into a pair of 
complex numbers $(z_1,z_2)$ as 
\begin{equation}
z_1 =p_1+i p_2 = \cos\frac{\theta}{2} e^{i \alpha}, \quad z_2 =p_3+i p_4=\sin\frac{\theta}{2} e^{i(\varphi+\alpha)}
\end{equation}
These complex coordinates are related to the spherical 
coordinates $\hat{\ps}(\theta,\varphi)$ on $S^2$ by the Hopf projection map $\pi: S^3 \rightarrow S^2$
\begin{eqnarray}
& & n_1 = z^{\ast}_1 z_2 +z^{\ast}_2 z_1 = \sin{\theta} \cos\varphi \nonumber \\
& & n_2 = i (z^{\ast}_2 z_1 - z^{\ast}_1 z_2) = \sin \theta \sin\varphi \nonumber \\
& & n_3 = |z_1|^2-|z_2|^2 = \cos\theta 
\label{Hopf123} 
\end{eqnarray}
where $(n_1,n_2,n_3)$ are the Cartesian coordinates of the unit wave-vector $\hat{\ps}$. 
Since locally the 3-sphere has a product form $S^2 \times S^1$, the Hopf projection has the 
property to eliminate the dependence on the third anHgle $\alpha$ by mapping 
the unit circle $S^1$ parameterized by $\alpha$ to a single point on $S^2(\theta,\varphi)$. 
The pair of complex coordinates can be grouped into a 
two-component spinor to label the points on $S^3$ 
\begin{equation}
z = \left( \begin{array}{c} 
z_1 \\
z_2
\end{array}
\right) = \left( \begin{array}{c}
\cos(\theta/2)e^{i \alpha} \\
\sin(\theta/2)e^{i(\varphi+\alpha)}
\end{array}
\right)
\end{equation}
In these coordinates the Hopf projection map in Eq.(\ref{Hopf123}) is written more simply 
\begin{equation}
\hat{\ps}(\theta,\varphi) = z^{\dagger} \boldsymbol\sigma z , 
\label{Hopf} 
\end{equation}
where $\boldsymbol\sigma=(\sigma_1,\sigma_2,\sigma_3)$ are the three Pauli matrices. 
A globally-defined connection one-form in these coordinates is given by
\begin{eqnarray}
& & \omega = -i z^{\dagger} d z = p_1 dp_2 - p_2 dp_1 + p_3 d p_4 - p_4 dp_3 = \nonumber \\
& & = d \alpha + \frac{1}{2} (1-\cos{\theta}) d \varphi 
\label{alpha}
\end{eqnarray}
which looks like the Wu-Yang potential one-form on the northern patch supplemented by an 
exact one-form $d \alpha$. The $S^3$ one-form is invariant under global $U(2)$ transformations
of the spinor coordinates, i.e. 
\begin{equation}
z \rightarrow U z, \quad U^{\dagger} U =1 , \quad \omega \rightarrow \omega 
\end{equation} 
The $SU(2)$ subgroup acts by the matrices 
\begin{equation}
U(\Omega,\ns) = \exp(i \Omega \ns \cdot \boldsymbol\sigma ), 
\end{equation}
and the result projected onto the two-sphere through the Hopf map as
\begin{equation}
z^{\dagger} (U^{\dagger} \boldsymbol\sigma U) z = \hat{\ps} \cos\Omega+ \hat{\ns}(\hat{\ns} \cdot \hat{\ps}) (1-\cos\Omega)
+ (\ns \times \hat{\ps}) \sin\Omega 
\end{equation}
generates rotation of the unit-vector $\hat{\ps}$ 
on an angle $\Omega$ about the vector $\ns$. Though the 
group $\{ A \in SL(2,\mathbb{C}) | {{\rm det A}}=1 \}$ of linear transformations of 
the spinor coordinates also acts on these states, 
the $S^3$ connection one-form $\omega$ is not invariant under such transformations. 
In orthogonal coordinates $(\theta,\xi=\varphi+\alpha, \alpha)$ 
that diagonalize the metric on the unit 3-sphere 
\begin{equation}
ds^2 = \frac{1}{4} d \theta^2 + \cos^2\frac{\theta}{2} d \alpha^2 + \sin^2\frac{\theta}{2} d \xi^2 
\end{equation}
the gauge-potential one-form in Eq.(\ref{alpha}) is 
\begin{equation}
\omega = \frac{1}{2} \omega_{\theta} d \theta + \cos\frac{\theta}{2} \omega_{\alpha} d \alpha + 
\sin\frac{\theta}{2} \omega_{\xi} d \xi , \label{omega3}
\end{equation}
non-singular with components 
\begin{equation}
\omega_{\theta}=0, \quad \omega_{\alpha}=\cos\frac{\theta}{2}, \quad \omega_{\xi}=\sin\frac{\theta}{2}
\end{equation}
The corresponding gauge-field strength two-form 
\begin{equation}
F = d \omega = -i z^{\dagger} \wedge dz = \frac{1}{2} \sin\theta d \theta \wedge d \varphi
\end{equation}
is exact and closed two-form $(dF=0)$ on $S^3$.  It gives half of the volume of the
two-dimensional unit sphere $S^2$ and therefore the topology of the 3-sphere can be 
characterized by the first Chern number 
\begin{equation}
c_1 =\frac{1}{2 \pi} \int_{S^2} F = +1, 
\end{equation}
which corresponds to a conserved helicity quantum number $\mu=+1/2$. 
The reduction to the Wu-Yang monopole gauge potentials over the 2-sphere can be obtained 
by taking sections of the 3-sphere. Local sections of $S^3$ can be taken if a particular value for 
the phase angle $\alpha$ is fixed, such that the metric on the 3-sphere 
reduces to the metric on the 2-sphere $ds^2 = d \theta^2 + \sin^2 \theta d \varphi^2$
in the north $R_N, (\theta/2 \rightarrow \theta)$ and south $R_S, (\pi-\theta)/2 \rightarrow \theta$ 
hemispheres, respectively. These two choices correspond to fixing $\alpha=0$ and $\alpha=-\varphi$ 
in Eq.( \ref{omega3}) and leading to the Wu-Yang monopole potentials on the two local patches 
\begin{equation}
A_N = \frac{1}{2} (1-\cos{\theta}) d \varphi ,
\quad A_S = -\frac{1}{2} (1+\cos{\theta}) d \varphi, 
\end{equation}
respectively. 
Using the property of the Hopf projection map, 
the regular $S^3$ connection one-form can be written in more compact way in terms of projection operators 
\begin{equation}
\omega = d \alpha z^{\dagger} \frac{1}{2} (1+\sigma_3) z + d \xi z^{\dagger} \frac{1}{2} (1-\sigma_3) z
\end{equation}
where $\sigma_3={{\rm diag}} (+1,-1)$ is the third Pauli matrix. The eigen-states of $\sigma_3$ 
\begin{equation}
\frac{1}{2} \sigma_3 | \sigma \rangle = \sigma | \sigma \rangle
\end{equation}
for $\sigma = \pm 1/2$ are two-component spinors 
\begin{equation}
(1,0)^T=|+\rangle, \quad =(0,1)^T = |-\rangle 
\end{equation}
which can represent the north $N=(0,0,1)$ and south poles $S=(0,0,-1)$ on $S^2$, 
as follows from the property of the Hopf map (cf. Eq.(\ref{Hopf})). 
In terms of these two states, the $S^3$ gauge-potential one-form 
\begin{equation}
\omega = d \alpha |\langle + | z \rangle |^2 + d \xi |\langle - | z \rangle |^2. 
\end{equation}
is written as a sum of separate probabilities for an $S^3$ point $z$ to have two different signatures $\pm$.  
For fixed $\theta$, the two circle variables  $(\xi,\alpha)$ parameterize a one-dimensional complex torus  
$T^2 = S^1 \times S^1 (e^{i \alpha}, e ^{i \xi}) $ in $S^3$, and hence locally $\omega$ is viewed as 
one-form over the torus. The  changes of $\theta$ generate
a family of tori $T^2(\theta)$. A linear equi-variation 
of the phase angles on a curve $\gamma(s)=(\alpha(s)=s,\xi(s)=s)$ gives 
\begin{equation}
\omega = ds, 
\end{equation}
i.e. $\omega$ reduces locally to an exact one-form. The unit Euclidean four-momentum changes as 
\begin{eqnarray}
& & n(s)=(\cos\frac{\theta}{2} \cos s, 
\cos\frac{\theta}{2} \sin s, \nonumber \\
& & \sin\frac{\theta}{2} \cos s, \sin\frac{\theta}{2} \sin s)
\end{eqnarray}
and describes a helix which lies in the flat torus $p_1^2 + p_2^2=\cos^2\theta/2, p_3^2 + p^2_4=\sin^2 \theta/2$. 
When the phase angle changes are not equi-variant $\alpha \ne \xi$, the helical paths the particle 
follows deform continuously. The two-component spinors $z(\theta,\xi,\alpha)$ can be used to 
represent a quantum state of additional spin-projection variable $\sigma$ that takes only 
two values $\sigma=\pm 1/2$. This is because, when the particle propagates north $\hat{\ps}=(0,0,1)$, 
and a local section onto the north hemisphere is taken 
\begin{equation}
A_N = d \varphi | \langle - | z \rangle |^2 = 0, \label{an}
\end{equation}
then $\sigma$ "points" north, since $| \langle + | z \rangle |^2=1$. 
Analogously, setting $\hat{\ps}=(0,0,-1)$ and taking southern local section
\begin{equation}
-A_S = d \varphi | \langle + | z \rangle |^2 = 0, \label{as}
\end{equation}
shows that the variable $\sigma$ is "pointing" south $| \langle - | z \rangle |^2=1$. 
In all cases, this can be written as 
\begin{equation}
A_{\sigma} = \sigma d \varphi (1-2 \sigma \cos \theta) 
\end{equation}
where the discrete variable $\sigma=\pm 1/2$ labels the two-coordinate patches. 
Therefore four coordinates $(p,\theta,\varphi,\sigma)$ can be used to parameterize 
the space, where the wave-function takes values. 
The invariant property of the gauge-field that correlates the spin and  momentum as 
given by Eqs.(\ref{an}) and (\ref{as}) can be made explicit, by noting that the 
sectional spin-states $\{z(\theta,\xi,\alpha), \alpha=0,-\varphi \}$
defined locally over the two-sphere, are eigen-states of the operator of the helicity, i.e. 
\begin{equation}
R_N: \frac{1}{2} \boldsymbol\sigma \cdot \hat{\ps} | z(\theta,\varphi,0) \rangle = +\frac{1}{2} |z(\theta,\varphi,0)\rangle 
\end{equation}
and 
\begin{equation}
R_S: \frac{1}{2} \boldsymbol\sigma \cdot \hat{\ps} | z(\theta,0,-\varphi) \rangle = +\frac{1}{2} |z(\theta,0,-\varphi)\rangle,  
\end{equation}
i,e, the helicity is conserved and can be identified with the first Chern number of 
the Hopf bundle $c_1/2=+1/2$. That is because the flux of the "background magnetic-like field" $\bc$ 
through the 2-sphere is the flux of the local spin vector field 
$\ssm(\theta,\varphi)=z^{\dagger} (\boldsymbol\sigma/2) z = +\hat{\ps}(\theta,\varphi)/2$ ((cf. also Eq.(\ref{spin}))
\begin{equation}
\frac{1}{2 \pi} \ointint_{S^2} \bc \cdot d\ssc= \int \int \frac{d \Omega }{2\pi} 
~\hat{\ps} \cdot \ssm(\theta,\varphi) =1 
\end{equation}
where $d \Omega=\sin\theta d \theta d \varphi$ is the area element on the two-sphere
and $\hat{\ps}$ is the outward surface normal. 
In terms of the sectional spin states $|\pm (\theta,\varphi) \rangle$, the spin-gauge fields 
on $S^2$ can be written as 
\begin{equation}
\ac(\ps,\sigma)=-\langle \sigma(\theta,\varphi) | i \nabla_{\ps} | \sigma(\theta,\varphi) \rangle 
\end{equation}
The angular momentum operator in Eq.(\ref{lcov}) takes the simpler form 
\begin{equation}
\lc = \rs \times \ps + \langle \sigma(\theta,\varphi) | \left( \rs \times \ps + \frac{1}{2} \boldsymbol\sigma \right) 
|\sigma(\theta,\varphi)\rangle, 
\end{equation}
and makes explicit the underlying total angular momentum operator folded between the  
states $|\pm(\theta,\varphi)\rangle$
\begin{equation}
{\bf J}=\rs \times \ps + \frac{1}{2} \boldsymbol\sigma 
\end{equation}
to be the conventional kinematic angular momentum $\ls = \rs \times \ps$ supplemented by a non-kinematic 
angular momentum operator $\frac{1}{2} \boldsymbol\sigma$ acting on the sectional spin states 
$\pm(\theta,\varphi)$. Therefore, the total gauge-invariant one-particle wave-function has an adiabatic 
form and representable by a product of orbital and spin-dependent factor 
\begin{equation}
\psi^{(\pm)}(\theta,\varphi) = Y^{(\pm)}_{l m \frac{1}{2}}(\theta,\varphi) |\pm(\theta,\varphi)\rangle  
\label{wigns}
\end{equation}
where $Y_{lm \mu}(\theta,\varphi)$ are the Wu-Yang wave-functions. 
The total wave-function is gauge-invariant, since 
when the $U(1)$ phase of the sectional spinor $|\pm(\theta,\varphi)$ is locally changed, the phase of the 
angular wave-function rotates oppositely, and the total wave-function remains gauge-invariant.

A dual Hopf bundle $H_{-1}(=S^3)$ corresponding to definite helicity quantum number $\mu=eg=-1/2$, or 
equivalently first Chern number $c_1=-1$ can be defined in terms of the 
conjugate left-handed spinors $\bar{z}=(-z^{\ast}_2,z^{\ast}_1)$, which satisfy $\boldsymbol\sigma \cdot 
\hat{\ps} \bar{z}(\theta,\varphi)=
-\bar{z}(\theta,\varphi)$.  Then the flux of the local spin vector field is 
\begin{equation}
\ointint \bc \cdot d \ssc = \int \int d \Omega ~\hat{\ps} \cdot \ssm(\theta,\varphi) = -2 \pi, 
\end{equation}
and that is why the flux of the "background magnetic-like field" has a negative sign, since the
spin-vector $\ssm$ points oppositely to the particle momentum $\ps$. Though the displacement operator 
$\nabla_{\ps}$ can couple states of different helicity $\mu=\pm 1/2$, 
these states exhibit opposite Chern character $c_1=+1$ and $c_1=-1$ and 
are topologically distinct. For free-particle states, 
the helicity $\mu = \pm 1/2$ is conserved. 

It can be pointed out, that the one-particle formalism can be generalized to a system of $N$ non-interacting 
particles, when the momentum space is $3N$ dimensional. The description involves a gauge-potential one-form 
\begin{equation}
A = d p^{\mu} A_{\mu}(p) 
\end{equation}
over $\mathbb{R}^{3 N}$, where $(\mu=1,2,\ldots, 3N)$ and $p=(p_1,p_2,\ldots , p_{3N})$ 
labels the points in $\mathbb{R}^{3N}$. However, different and more complex 
cooperative effects occur, since $A_{\mu}$ correlates the actions of single-particle 
angular momenta operators in non-trivial way. In particular, this cooperative effect, 
could describe the effect of particle inter-change and statistics.

\section{Screening } 
When there is an external potential field $U(\rs)$ present, the single-particle Hamiltonian is 
\begin{equation}
H = \frac{1}{2} \ps^2 + U(\rs) 
\end{equation}
and the corresponding Schr\"{o}dinger equation for the eigen-states is 
\begin{equation}
H | \Psi\rangle = E |\Psi\rangle , \label{eigen}
\end{equation}
By assuming that helicity is conserved adiabatically, 
i.e. assume that a selection-rule $\Delta \mu= 0$ is satisfied. 
That is because the external field $U(\rs)$ is assumed to be topologically 
trivial, such that it can not change the helicity of the particle $\mu$. 
On each patch, the total spin wave-function exhibits an adiabatic product form 
\begin{equation}
\Psi(\ps) = \psi(\ps) |z(\theta,\varphi)\rangle 
\label{state} 
\end{equation}
and that is why the patch label is suppressed. 
The orbital part of the wave-function can be expanded over complete set of 
monopole wave-functions 
\begin{equation}
\psi(\ps) = \sum_{lm} F_{lm \mu}(p) Y_{lm \mu} (\theta,\varphi) 
\end{equation}
where $\mu=eg$ is the helicity, $l=|\mu|,|\mu|+1,\ldots$ 
is the  orbital angular momentum quantum number and $m=-l, \ldots, l$ is the azimuthal 
quantum number. Expanding the external potential over Fourier components 
\begin{equation}
U(\rs) =\sum_{\qs} U(\qs) e^{i \qs \cdot \rs} 
\end{equation}
substituting Eq.(\ref{state}) into Eq.(\ref{eigen}), and 
projecting the result onto a helicity eigenstate, we obtain the equation for the 
orbital part of the wave-function
\begin{equation}
(\ps^2/2-E)\psi_E(\ps) + \sum_{\qs} U(\qs) F_{\ps}(\qs) \psi_E(\ps-\qs) =0, 
\end{equation}
and a form-factor has been introduced 
\begin{equation}
F_{\ps}(\qs) = \langle z(\ps) | e^{i \qs \cdot \rs} | z(\ps) \rangle =\langle z(\ps) | z(\ps-\qs) \rangle, 
\end{equation}
which has the effect to screen the Fourier components of the external potential 
$U(\qs) \rightarrow U_{{\rm eff}}(\ps,\qs)=U(\qs) F_{\ps}(\qs)$. The 
spin-dependent form-factor can be computed from the function 
\begin{equation}
F(t) = \langle z(\ps) | z(\ps(t)) \rangle \label{fft}
\end{equation}
on a straight line path $\ps(t) = \ps(1-t) + (\ps-\qs) t, (0 \le t \le 1)$ inter-connecting the
wave-vectors $\ps$ and $\qs$. The function $F(t)$ satisfies the initial condition $F(t=0)=1$ 
and $F(t=1)=F_{\ps}(\qs)$ gives the form-factor. Differentiating Eq.(\ref{fft}) 
with respect to the parameter $t$ gives 
\begin{equation}
\dot{F}(t) = \dot{\ps} \cdot  \langle z(\ps) | \nabla_{\ps} | z(\ps(t)) \rangle 
\end{equation}
and using that $|z(\ps(t))\rangle = F(t) |z(\ps(0)) \rangle$, an 
equation for the Berry's phase is obtained 
\begin{equation}
\dot{F}(t)= \frac{1}{i} \dot{\ps} \cdot \ac(\ps(t))  F(t), \label{eomh}
\end{equation} 
where $\ac(\ps)=\langle z(\ps) | i \nabla_{\ps} | z(\ps) \rangle$ is the gauge field. Eq.(\ref{eomh}) 
can be integrated along the straight line to give at the end point $t=1$ the form-factor 
\begin{equation}
F(t=1) = F_{\qs}(\ps) = \exp \left( i \int_{\ps-\qs}^{\ps} d \ks \cdot \ac(\ks) \right). 
\end{equation}
The Schr\"{o}dinger equation reduces to a pair of coupled equations for the 
wave-functions on the patches 
\begin{eqnarray}
& & (\ps^2/2 -E) \psi_N(\ps) + \sum_{\qs \in R_N} U(\ps-\qs) F_N(\ps,\qs) \psi_N(\qs) + \nonumber \\
& & + \sum_{\qs \in R_S} U(\ps-\qs) F_{NS}(\ps,\qs) \psi_S(\qs)=0 \nonumber \\
& & (\ps^2/2 -E) \psi_S(\ps) + \sum_{\qs \in R_S} U(\ps-\qs) F_S(\ps,\qs) \psi_S(\qs) + \nonumber \\
& & + \sum_{\qs \in R_N} U(\ps-\qs) F_{SN}(\ps,\qs) \psi_N(\qs)=0
\label{patch-eom}
\end{eqnarray}
where the patch label $N(S)$, instead of $\sigma=\pm 1/2$, is used. 
The  form-factor for the northern patch is
\begin{equation}
F_N(\ps,\qs)=\exp \left( i \int_{\qs}^{\ps} d \ks \cdot \ac_N(\ks) \right), 
\end{equation}
and similarly on the southern patch it is 
\begin{equation}
F_S(\ps,\qs)=\exp \left( i \int_{\qs}^{\ps} d \ks \cdot \ac_S(\ks) \right). 
\end{equation}
These form-factors can be evaluated from the overlap of the 
sectional spinors $\langle \pm (\qs) | \pm (\ps) \rangle$, for instance 
\begin{equation}
F_N(\qs,\ps)= \cos\frac{\theta}{2} \cos\frac{\theta'}{2}+ \sin\frac{\theta}{2} 
\sin\frac{\theta'}{2}e^{i (\varphi-\varphi')} . 
\end{equation}
where $\qs=(q,\theta',\varphi')$ and $\ps=(p,\theta,\varphi)$. 
When the integration path crosses the equator, the spin projection $\sigma$ onto the $z$-axis flips, 
i.e. the coordinate patches interchange, and the phase-factor is evaluated from 
(cf. also Ref. \cite{wu-yang}) 
\begin{equation}
F_{NS}(\ps,\qs)=F_N(\ps,\ks_E) e^{i \varphi_{NS}(\ps,\qs)} F_S(\ks_E, \qs) 
\end{equation}
Here $\ks_E$ is an equatorial vector in the overlap region of the two patches. It is 
determined from the crossing of the straight line interconnecting the pair $(\qs,\ps)$ 
with the equatorial plane $(\theta=\pi/2)$. 
Explicitly $\ks_E$ is given by 
\begin{equation}
\ks_E = -\qs \frac{p_z}{p_z-q_z} + \ps \frac{q_z}{p_z-q_z}  = \frac{(\ps \times \qs) \times \ns }
{\ps \cdot \ns - \qs \cdot \ns}
\end{equation}
where $\ns=(0,0,1)$ is the unit-vector pointing along the $z$-axis. The azimuthal 
angle $\varphi_{NS}$ is given by 
\begin{equation}
{{\rm tan}} \varphi_{NS}(\ps,\qs)= \frac{k_y}{k_x} = \frac{p_y q_z - q_y p_z}{p_x q_z - q_x p_z} 
\end{equation}
The spin-transition form-factor $F_{SN}$ 
can be obtained from symmetry relation $F_{SN}(\ps,\qs)=F^{\ast}_{NS}(\qs,\ps)$. 
The matrix elements $U_{l'm',lm}(p,q)$ of the screened potential are evaluated in basis of monopole 
harmonics $Y_{lm \mu}(\theta,\varphi)$ and coupled integral equations 
for the partial-wave amplitudes $F_{lm \mu}(p)$ must be integrated numerically. 
This mathematical formalism can be applied to the case of simple 
harmonic oscillator potential $U(r)=r^2/2$, 
when numerical computation is not needed. The effective Hamiltonian for this particular case is 
\begin{equation}
H_{{\rm eff}} = \frac{1}{2} (i \nabla_{\ps} - \ac(\ps))^2 + \frac{1}{2} p^2 
\end{equation}
and can be diagonalized in the basis of the spin-weighted Wu-Yang monopole harmonics, i.e. 
\begin{equation}
\psi(\ps)=Y_{lm \pm \mu}(\theta,\varphi) F_l (p) . 
\end{equation}
where $\mu=\pm 1/2$ is the helicity,  $l=1/2,3/2,\ldots$ is the orbital angular momentum quantum number 
and $m=-l,-l+1,\ldots, l$. The effective Hamiltonian for radial motion is 
\begin{equation}
H_l(p) = -\frac{1}{2p} \frac{d^2}{d p^2} p + \frac{l(l+1)-1/4}{2 p^2} + \frac{1}{2} p^2, \label{hlp}
\end{equation}
i.e. the gauge-field only changes the effective centrifugal barrier for radial motion. 
The wave-functions of Eq.(\ref{hlp}) are analytic and given by means of generalized 
Laguerre polynomials 
\begin{equation}
F_{vl}(p) = p^{l^{\ast}} e^{-p^2/2} L_v^{l^{\ast}+1/2}(p^2), 
\end{equation}
where $l^{\ast}=\sqrt{l(l+1)}-1/2$ is an effective fractional angular momentum and $v=0,1,2,\ldots$ 
is a vibrational quantum number, which counts the nodes of the momentum-space wave-functions. 
The energy levels of the helicity-carrying oscillator eigen-states 
are $l$-dependent 
\begin{equation}
E_{vl}=2 v + \sqrt{l (l+1) } + 1
\end{equation}
and for a given $v$, levels are $(2l+1)$-fold degenerate, corresponding to their 
independence on the magnetic quantum number $m$. 
In the simplest case, when the particle is spin-less $\mu=eg=0$, 
the harmonic oscillator energy levels are given by 
$2v + l_0 +3/2$ for $l_0=0,1,2,\ldots$. These states can be labeled by a single quantum number $N=2v+l_0$. 
Each level of principal quantum number $N$ is $(N+1)(N+2)/2$-fold degenerate. In opposite, when the particle 
is carrying a helicity $\mu=\pm 1/2$, this degeneracy is lifted, and states can 
not be classified by a single quantum number $N$. This is because the effective 
orbital angular momentum $l^{\ast}=\sqrt{l(l+1)}-1/2$ is fractional. 
Therefore the principal effect of the gauge-field is to lift degeneracy of 
conventional harmonic oscillator energy levels, which split depending 
on both the vibration quantum number $v=0,1,2,\ldots$ 
and the orbital angular momentum quantum number $l=1/2,3/2,\ldots$. 

In the case of hydrogen atom, represented by a Coulomb potential $Z r^{-1}$, 
similar effect of splitting of the energy levels can occur due to 
the screening of the Coulomb field by the helicity-carrying particle. 
We expect that the effect of energy-level splitting is small, as 
the observed fine-structure of energy levels of hydrogen shows, 
and numerical computation must be made in order to verify 
if such an effect of spin-dependent screening is small or negligible.

\section{Conclusion} 
The local phase invariance of the momentum-space Schr\"{o}dinger equation has been used 
to describe the motion of a non-relativistic particle with spin and helicity. 
As a byproduct, effective one-particle Schr\"{o}dinger equation of motion 
in external field is derived, which predicts an effect of spin-dependent screening of the external 
potential. The approach is applied for simple harmonic oscillator potential and shown that 
the effect of screening affects rotation energy level splittings.


\begin{thebibliography}{99}
\bibitem{Wilczek} F.~Wilczek, A.~Zee,  Phys.~Rev.~Lett. {\bf 52}, 2111 (1984).
\bibitem{Berry} M.~V.~Berry, Proc.~R.~Soc.~Lond. A {\bf 392}, 45 (1984).
\bibitem{Wilczek-book} A.~Shapere and F.~Wilczek, \emph{Geometric
phases in physics} (World Scientific, Singapore, 1989).
\bibitem{fiber} A.~Tomita  and R.~Y.~Chiao, Phys.~Rev.~Lett. {\bf 57}, 937 (1986). 
\bibitem{NQR} R.~Tycko, Phys.~Rev.~Lett. {\bf 58}, 2281 (1987). 
\bibitem{Na3} G.~Delacr\'etaz, E.~Grant, R.~Whetten, L.~W\"oste, J.~W.~Zwanziger, 
Phys.~Rev.~Lett. {\bf 56}, 2598 (1986).
\bibitem{Hall1} Z.~Fang, N.~Nagaosa, K.~Takahashi, A.~Asamitsu, R.~Mathieu, T.~Ogasawara, 
H.~Yamada, M.~Kawasaki, Y.~Tokura, K.~Terakura, Science {\bf 3}, Vol. 302, no. 5642, pp. 92-95 (2003). 
\bibitem{Hall2} Y. Zhang, Yan-Wen Tan, Horst L. Stormer,  Philip Kim
Nature {\bf 438}, 201-204 (2005). 
\bibitem{Jackiw3} R.~Jackiw, Int.~ J. Mod. Phys. {\bf A, 3}, pp. 285-297 (1988).
\bibitem{Jackiw2} R.~Jackiw, Phys. Rev. Lett. {\bf 54}, 159 (1985). 
\bibitem{monopole-book} Ya.~Shnir, \emph{Magnetic monopoles} (Springer-Verlag, Berlin, Heidelberg, 2005).
\bibitem{Aitch} I.~J.~R.~Aicthison, Acta Physica Polonica, {\bf B 18}, p.207 (1987). 
\bibitem{wu-yang} T.~T.~Wu and C.~N.~Yang, Phys. Rev. {\bf D} 12, 3845 (1975). 
\end{thebibliography}
\end{document}